# El Sonido como Elemento Clave en Prácticas de Realidad Virtual

Yesid Ospitia Medina
Facultad de Informática
Universidad Nacional de la Plata, UNLP
La Plata, Argentina
yesid.ospitia@gmail.com

***Resumen.*** En este artículo se estudia la importancia del sonido para los sistemas de realidad virtual (RV). Para ello, se analiza la afectación emocional generada por el sonido, y su contribución al efecto de inmersión de los sistemas de RV.

***Palabras claves—*** *RV (realidad virtual), inmersión, TI (Tecnologías de la Información), CA (Computación afectiva)*

***Abstract—*** This article discusses the importance of sound for virtual reality systems. For this, the emotional effects generated by sound are analyzed, and its contribution to the effect of immersion.

## I. Introducción

La realidad virtual en su definición más general, puede ser considerada como un medio de interacción entre el ser humano y un mundo artificial.

La calidad y precisión de los diferentes canales sensoriales son parte clave del éxito de los sistemas de realidad virtual. Estos canales y su desempeño en tiempo real son los responsables de generar el mejor efecto de inmersión.

La inmersión es uno de los conceptos fundamentales en los sistemas de realidad virtual, cuyo objetivo principal, es generar una experiencia lo más real posible. Los sistemas de realidad virtual necesitan considerar todo tipo de condiciones de contexto, para simular experiencias con el mayor grado de inmersión. De manera general, la realidad virtual debe trabajar 3 aspectos que se encuentran presentes en la interacción con el mundo real: la visión, el sonido y el tacto.

Lo que se espera de un sistema de realidad virtual altamente efectivo, es que su efecto de inmersión sea lo más convincente posible. Y esto se logra cuando una persona reacciona ante un sistema de realidad virtual (RV) tal como reaccionaría ante un mundo real.

La forma en que una persona reacciona se refleja generalmente a través de sus emociones. Por lo que un buen indicador de efectividad sobre un sistema de RV, puede ser la medición de la intensidad de las emociones. Si un sistema de RV logra generar las mismas emociones presentadas en una experiencia real, podemos afirmar que su objetivo se ha cumplido en gran medida.

El estudio de las emociones se ha considerado un gran desafío para la ciencia. Tradicionalmente la ciencia, parte de argumentos lógicos y de hipótesis comprobables a través de experimentos que permiten explicar diferentes fenómenos. Pero este esquema tradicional de razonamiento, no ha sido suficiente para explicar con precisión las emociones.

Las emociones como objeto de investigación han generado frustración, debido a la dificultad de entendimiento desde la perspectiva científica tradicional. Pero sin duda alguna es un campo de gran interés, si reconocemos la importante relación existente entre las emociones y el bienestar general de un ser humano. [1, p. 1]

Desde la psicología cognitiva podemos afirmar que las emociones que sentimos son producto de nuestros pensamientos. Los pensamientos se originan de las evaluaciones que establecemos sobre hechos reales. El juicio que damos sobre un hecho en particular varía mucho dependiendo de nuestra escala de evaluación y de la rigurosidad con la cual afrontamos la vida. Finalmente, la emoción es la materialización de dicha evaluación. [2]

El estudio de las emociones nos resulta importante, porque nos permite entender lo que está sucediendo con una persona, mientras experimenta una determinada acción. Para el campo particular de tecnologías de la información (TI), nos permite construir sistemas que puedan reconocer emociones, evaluarlas y responder adecuadamente; mejorando de esa manera la posible frustración que muchos usuarios pueden presentar al momento de trabajar con diversas tecnologías.

La computación afectiva como campo de investigación emergente, se concentra específicamente en el entendimiento de emociones y su relación con las TI. Cabe aclarar que para mejorar la comunicación con el usuario, los periféricos tradicionales han evolucionado y también han aparecido nuevos dispositivos de entrada de datos. Todo obedece a que un teclado y un ratón no es suficiente para determinar con precisión las emociones, y entender la experiencia del usuario al trabajar con una máquina.

En la actualidad existe una amplia variedad de dispositivos que permiten capturar señales generadas por el usuario y procesarlas. A partir del procesamiento de esta información se puede establecer una gran cantidad de aplicaciones para mejorar la experiencia del usuario en su interacción con las TI. Parte del reto de la computación afectiva (CA) se encuentra en los requerimientos de hardware y en los posibles altos costos de los periféricos, debido a que la precisión de medición es un requerimiento fundamental. A mayor precisión y calidad del instrumento, generalmente tenemos un mayor costo de adquisición.

Las expresiones faciales son uno de los medios más explorados y reconocidos dentro del campo de la computación afectiva. Duchenne de Boulonge en su tesis de 1862, desarrolló un estudio sobre el comportamiento de los músculos de un rostro [1, p. 6]; aportando importantes fundamentos para modelar expresiones faciales a través de la computación, y dando los primeros elementos, para luego vincular la investigación a la psicología.

Las expresiones faciales son analizadas por la computación afectiva, y permite determinar que emociones son experimentadas por un usuario a partir de una vivencia. La experiencia del usuario esta condicionada a una gran cantidad de variables de contexto, el cual se encuentra determinado en gran medida por lo que se observa y lo que se escucha.

El propósito de este trabajo es estudiar el impacto generado por la música, como una posible condición de contexto que experimenta un usuario al interactuar con un sistema de realidad virtual. Para ello se debe realizar un ejercicio, que permita integrar contenido emocional, computacional y musical.

Debemos aclarar que sonido no es lo mismo que música. El sonido puede ser considerado un concepto de muy alto nivel, con argumentación física. Pero, no todo sonido trasciende al ámbito emocional. El ruido de una puerta, de un equipo electrónico, o de una marcha protestante por ejemplo, generalmente no transmite contenido emocional importante. La música tiene unas características fundamentales, que a lo largo de la historia han constituido unas reglas básicas para su propio desarrollo. La música es un lenguaje, y se requiere entender sus principios fundamentales, para abordar su relación con las emociones. [3, p. 7]

La música tiene un efecto directo sobre las emociones y el estado de ánimo de las personas. Puede ser considerada como un elemento de transformación emocional. De manera general y a través de conocimiento empírico, una gran cantidad de oyentes manifiestan sentir cambios emocionales al escuchar la música. [4]

Muchas personas acostumbran escuchar cierto tipo de música para combatir el estrés, e incluso otros sentimientos profundos. Existen canciones que contribuyen a superar la frustración, a olvidar momentos desagradables, o incluso a motivar y movilizar grandes ejércitos para enfrentar una batalla. Las razones exactas que expliquen este vínculo entre las emociones y la música, son objeto de estudio; y es parte de lo que se abordará en este artículo, como un laboratorio práctico cuyos resultados nos permitirán aproximarnos a establecer un vínculo entre la música y las emociones, llamando así la atención sobre su importancia para facilitar la inmersión en la realidad virtual.

Pensemos en un ejercicio sencillo, que consiste en escuchar 2 piezas de música clásica muy famosas. *Lacrimosa (Requiem)* de Mozart y *el cuarto movimiento de la novena sinfonía* de Bethoveen, comúnmente conocido como el *Himno a la Alegría*. Dos piezas musicales de gran complejidad y elaboración musical, pero quizás lo más importante, es que tienen efectos emocionales totalmente diferentes.

Ante tal ejercicio se plantean las siguientes preguntas:

- ¿Qué siente el oyente al escuchar una determinada canción? .
- ¿Qué puede imaginarse el oyente?.
- ¿Por qué el compositor elaboró esa música?.
- ¿Cuál fue la inspiración del compositor?.
- ¿Qué es la inspiración y donde tiene origen?.
- ¿Porqué reflejamos una emoción diferente cuando escuchamos cada pieza musical?.
- ¿Cómo afecta cada canción al oyente durante la experiencia de un mismo escenario de realidad virtual?

Entre ambas canciones existe un contraste diferenciador que de manera general lleva al oyente a experimentar cosas diferentes. ¿Cuál de las obras puede incrementar el estrés? ¿Cuál de las obras disminuye el estrés en el oyente? Y, ¿Por qué se da este efecto?

*Lacrimosa* forma parte de una obra famosa conocida como *La Misa de Réquiem*. Esta pieza musical fue compuesta por Wolfgang Amadeus Mozart y era interpretada en actos litúrgicos tras la muerte de una persona. Una obra musical para un contexto que en la mayoría de las culturas genera tristeza.

Es difícil afirmar que existe un criterio universal para percibir emociones desde la música. De hecho no todos los oyentes son amantes de la música clásica. La música que escuchamos la elegimos con base en una gran cantidad de variables de contexto, lo que afecta directamente el objeto de estudio de este artículo.

En este artículo se revisará específicamente el sonido como uno de los elementos claves para prácticas de realidad virtual. Para ello, se ejecutó un laboratorio en donde a través de análisis de rasgos faciales, fueron estudiadas las reacciones emocionales, a partir de estimulaciones sonoras. Adicionalmente, se utilizó una evaluación (test) de experiencia de usuario para obtener mayor información del ejercicio, como también una entrevista con cada uno de los 5 voluntarios.

Finalmente se ejecutó un laboratorio complementario, en donde se incluyó un escenario de realidad virtual, para conocer el efecto de la estimulación sonora sobre un escenario de juego, y así obtener retroalimentación a partir de la experiencia de usuario.

## II. Herramientas de reconocimiento facial

Se describe a continuación las herramientas que son objeto de estudio en este artículo, antes de iniciar con el primer laboratorio.

### A. Google cloud plataform Vision API[5]

Este *API (Application Programming Interface)* de Google permite a los desarrolladores procesar e interpretar una imagen a través de modelos de aprendizaje automáticos. De manera muy rápida el *API* facilita la clasificación de imágenes, como también la detección de objetos individuales, y caras dentro de una imagen.

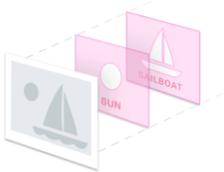

**Fig 1:** Análisis de Imágenes.
**Fuente:** Google Cloud Platform Web Site

El software también permite detectar contenido inapropiado. Para ello hace uso de los filtros definidos por *SafeSearch* de google [6]. Estando en capacidad de detectar contenido sexual explícito.

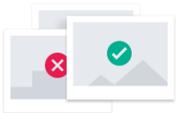

**Fig 2:** Detección de imágenes inapropiadas.
**Fuente:** Google Cloud Platform Web Site

El análisis de emociones en las imágenes es otra de las características importantes de este *API*. Adicionalmente, es posible identificar un objeto dentro de la imagen y procesar la expresión facial para determinar el impacto de este objeto en la persona. Algo muy útil para evaluar el contenido emocional de un logo.

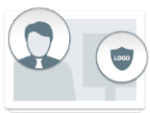

**Fig 3:** Análisis de Imágenes.
**Fuente:** Google Cloud Platform Web Site

El *OCR (Optical character recognition)* es otra de las funcionalidades destacables en este producto. Facilitando la detección de texto con reconocimiento automático del idioma.

### B. Open Face[7]

*Open Face* como herramienta de reconocimiento facial inició su desarrollo en la Universidad de Carnegie Mellon de Pittsburgh. La herramienta se encuentra desarrollada en lenguaje *Python* y adicionalmente hace uso de un *framework* conocido como *Torch*, el cual implementa algoritmos de aprendizaje automático. Una característica importante es su código abierto y de libre uso.

La herramienta permite detectar una cara, transformarla, recortarla y finalmente representarla computacionalmente. Entre otras funcionalidades interesantes, permite clasificar las imágenes, realizar detección de similitudes, y generar agrupaciones por criterios definidos.

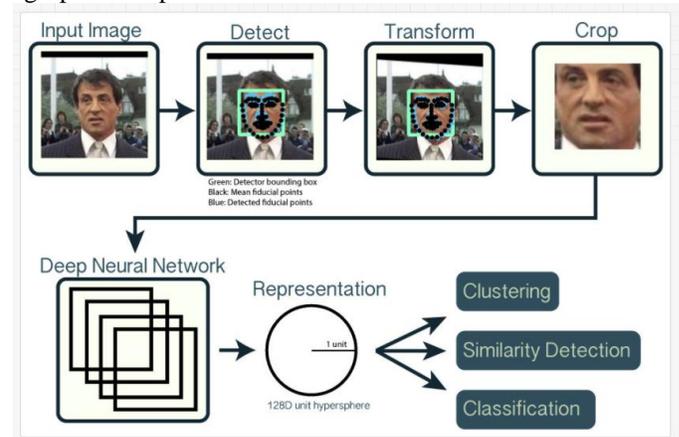

**Fig 4:** Esquema de funcionamiento de Open Face.
**Fuente:** Open Face Web Site

En pruebas realizadas con personajes reconocidos de la farándula, la herramienta ha mostrado una efectividad del 87% de precisión ejecutando una acción de reconocimiento facial.

*Open Face* puede ser un elemento fundamental para diversos proyectos de propósito académico. El licenciamiento definido por este proyecto facilita recoger sus avances y seguir construyendo nuevas funcionalidades.

Por otra parte también debemos considerar que el proyecto es bastante nuevo y por lo tanto aún falta mucho para madurar sus características. Directamente en el sitio de *github* [8], en donde se publican las versiones de la herramienta, se puede evidenciar el trabajo realizado durante los últimos 2 años. La primer versión (RELEASE) de la herramienta, fue lanzada el 13 de octubre de 2015.

En el sitio web en donde se manejan las versiones publicadas de la herramienta, se reportan varias fallas (ISSUES) que nos permiten especular un poco sobre la estabilidad y la perspectiva de usuario.

Algunos de los casos reportados están relacionados con problemas técnicos de instalación y con la falta de una mejor documentación.

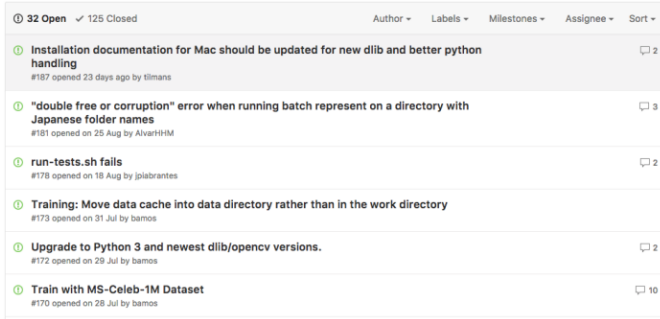

**Fig 5:** Issues de Open Face.
**Fuente:** Open Face Web Site GIT HUB

*C. Affdex*[9]

*Affectiva* es una compañía pionera en el desarrollo de sistemas de reconocimiento facial, la experiencia y el desarrollo se hace evidente en las 7 patentes que se le han otorgado hasta el momento y un poco de más de 30 casos de patentes en estudio de aprobación.

*Affdex* es una de las herramientas más conocidas en el campo de reconocimiento facial y emocional. Sus primeros pasos se dieron en investigaciones desarrolladas en el laboratorio de multimedia del MIT.

Entre las características fundamentales podemos encontrar las siguientes:
- Procesamiento de imágenes.
- Detección de rostros.
- Identificación de emociones.
- En los últimos años *Affdex* ha procesado más de 1 billón de *frames* alrededor de 75 países. A la fecha cuenta con un repositorio de información emocional que supera los 3.9 millones de rostros. Esta característica es en una de las grandes virtudes de *Affdex*.
- Interpretación de emociones en tiempo real. Lo que fortalece la confiabilidad de los resultados. Esta característica en *Affdex* la definen como optimización.

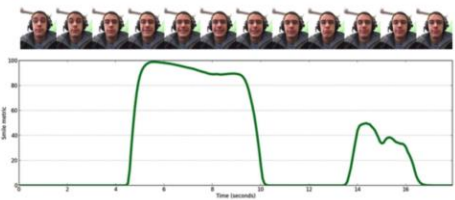

**Fig 6:** Utilización de métricas para analizar y mejorar la confiabilidad en reconocimiento de emociones.
**Fuente:** Affdex Web Site.

- Posibilidad de realizar predicciones basado en tableros de control visuales y de fácil interpretación.
- En la fase de evaluación *Affdex* dispone de un módulo que permite clasificar y desplegar los resultados por criterios como geografía y otras categorías.
- Posibilidad de analizar múltiples rostros en una misma imagen.
- Capacidad de reconocer un rostro con presencia de gafas, y también de identificar el género.

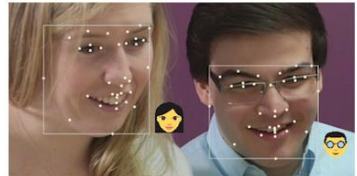

**Fig 7:** Identificación de género y detección de gafas.
**Fuente:** Affdex Web Site.

- Consideración de posición de cabeza y orientación de la luz.

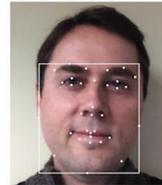

**Fig 8:** Seguimiento a la posición del rostro.
**Fuente:** Affdex Web Site

- Algoritmos de reconocimiento desarrollados con alta calidad y verificados con rigurosas pruebas. Lo que fortalece la confiabilidad en el reconocimiento de emociones.
- Compatibilidad en multiplataforma, alojamiento en la nube y compatibilidad con dispositivos móviles.
- Análisis de respuestas emocionales frente a marcas comerciales.

*D. Noldus Face Reader*[10]

*Noldus* como compañía se ha caracterizado por participar en comunidades científicas promoviendo la investigación y la Las investigaciones de *Noldus Company* sobre comportamientos, permiten establecer decisiones en un contexto determinado.

*Noldus Company* fue fundada en 1989, y en principio sus investigaciones estuvieron concentradas en el comportamiento ecológico y animal. El éxito de estos primeros proyectos, le permitieron a *Noldus Company* afrontar nuevos retos y desarrollar nuevas herramientas.

Uno de los aspectos a resaltar de *Noldus Company*, es su importante compromiso con lineamientos éticos. Todo en consideración a que *Noldus Company* realiza algunos procedimientos con animales para formular diferentes tipos de medicinas.

*FaceReader* es una de las aplicaciones de reconocimiento facial desarrollada por *Noldus Company*.
Entre las principales características de la herramienta tenemos:

- Reconocimiento y análisis de expresiones faciales.
- Repositorio con más de 10.000 imágenes que son utilizadas para el entrenamiento del software en el proceso de reconocimiento facial.
- Modelado preciso de caras con 500 puntos clave.

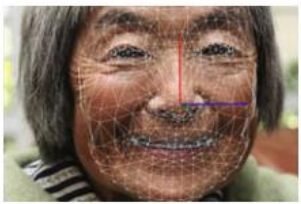

**Fig 9:** Modelado preciso de la cara.
**Fuente:** Noldus Face Reader Web Site.

- Disponibilidad de un API para integrar la tecnología con otras herramientas.
- Consideración de la dirección de la mirada y orientación de la cabeza.
- Analizar cómo las personas responden ante nuevos diseños de comerciales.
- Identificación del género y determinación de la edad.
- Dentro de las novedades de *facereader* encontramos un tablero de indicadores muy preciso, que permite analizar los cambios emocionales.

*A. Emotion API and Face Detetion*[11]
Microsoft ha trabajado en un API de reconocimiento facial e identificación de emociones, que permite detectar un rango de expresiones y con base en ello formular respuestas desde las aplicaciones de software.

innovación. Lucas Noldus, con un doctorado en comportamiento animal, es el fundador y director de la empresa. El principal objetivo de *Noldus Company* es la medición y el análisis de todo tipo de comportamientos.

Dentro de las características principales tenemos:

- Posibilidad de identificar caras individuales dentro de una misma foto y ejecutar su respectivo análisis, determinando para cada cara la emoción expresada.

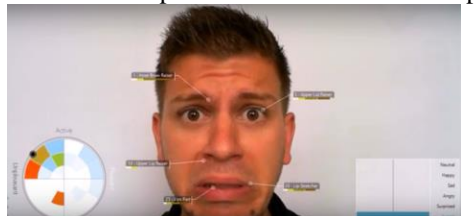

**Fig 10:** Modelado preciso de la cara .
**Fuente:** Microsoft Emotion API Web Site

Además de realizar el reconocimiento directamente en una imagen puede realizarlo en tiempo real a través de un video.

- La detección de caras se logra a través del módulo *face detection*. Dentro de sus principales características encontramos:
    - Verificación de rostros, para determinar si en 2 fotos independientes se encuentra presente la misma persona.
    - Medición remota del ritmo cardiaco a través del análisis de pequeños cambios de color generados en la piel.
    - Agrupamiento de caras en grupos con similitud visual.

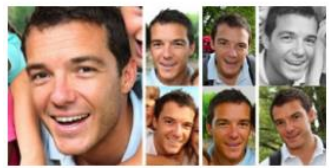

**Fig 11:** Agrupamiento de caras.
**Fuente:** Microsoft Emotion API Web Site.

- Realizar búsqueda de caras con atributos de similitud.

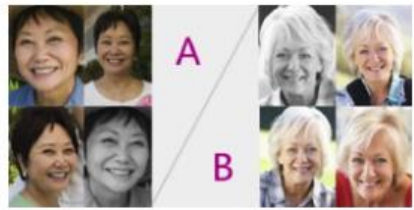

**Fig 12:** Búsqueda por similitud de cara.
**Fuente:** Microsoft Emotion API Web Site

## III. COMPARATIVO RECONOCEDORES FACIALES

A continuación se muestra una tabla comparativa de los atributos evaluados para las 5 herramientas estudiadas en este artículo. La "X" en la tabla representa que el producto cuenta con el atributo. En algunos casos la X se encuentra acompañada de un número entre 1 y 3. Para lo cuál se debe tener en cuenta la siguiente escala: 1 – Deficiente, 2 – Aceptable, 3 – Destacado. En este último caso, además de verificar la presencia del atributo en la herramienta, se asigna una evaluación dependiendo de la calidad de la presencia del atributo en el producto. Le evaluación fue realizada considerando la información de los sitios web oficiales de cada herramienta.

| Atributo / Producto | Cloud Vision API | Open Face | Affdex | Facer Reader Noldus | Emotion API - Microsoft |
|---|---|---|---|---|---|
| Detección de varios objetos en una misma imagen | X | | X | X | X |
| Clasificación de imágenes en categorías | X | X | X | | X |
| Detección de contenido inapropiado | X | | X | | |
| Detección y análisis de atributos emocionales | X | X | X | X | X |
| OCR | X | | X | | |
| Detección de logo de productos | X | | X | | |
| Código Abierto | | X | X | | |
| Libre Uso (Free) | | X | X | | |
| Usabilidad (indicadores, gráficos, tableros de control) – Fácil interpretación | X (2) | | X (3) | X (3) | X (2) |
| Disponibilidad de APIs para integrar con otras aplicaciones | X | X | X | X | X |
| Funcionamiento como un servicio en la nube | X | | X | X | X |
| Repositorio de rostros, emociones, etnia, otros atributos. | | | X (3) | X (2) | X (2) |
| Identificación de Género | | | X | X | X |
| Consideraciones éticas | | | | X | |
| Consideración en comportamientos de animales | | | | X | |
| Calcular la edad. | | | | X | X |
| Medición remota del ritmo cardiaco | | | | X | |
| Análisis de emociones en tiempo real - Video | X | X | X | X | X |
| Búsqueda de rostros similares. | | | | | X |
| Agrupamiento de caras por similud de atributos faciales. | | | X | | X |
| Calidad y detalle de la documentación | (X) 3 | (X) 1 | (X) 3 | (X) 2 | (X) 2 |

## IV. CONCLUSIONES RELATIVAS AL ANÁLISIS DE RECONOCEDORES FACIALES

- El estudio de las emociones continúa siendo un reto para la ciencia.
    - Por una parte se requiere seguir trabajando en la comprensión de las emociones desde un punto de vista psicológico, y generar un esfuerzo para establecer métricas con criterios objetivos; lo que resulta sumamente complejo, porque debemos involucrar la mayor cantidad de señales y determinar la confiabilidad de estas mismas.
    - El desarrollo de las interfaces de hardware para mejorar la comunicación humano / máquina es otro aspecto de gran complejidad. Desarrollar el hardware es una condición necesaria para mejorar las condiciones de investigación e incrementar la precisión de los diversos estudios. La complejidad de diseño y los costos son otro aspecto importante a manejar desde la computación afectiva.
- Los avances de la computación afectiva en cuanto a desarrollos sobre reconocimiento facial son considerables. Se puede evidenciar una serie de herramientas existentes con diversas características. En su mayoría el enfoque se concentra en el reconocimiento facial y la clasificación de emociones.
- Como resultado del análisis comparativo entre las herramientas, encontramos que todas parten de un principio de funcionamiento; el reconocimiento facial. A partir de esa funcionalidad principal, las diferentes instituciones crean funcionalidades complementarias. Las funciones complementarias fueron descritas en la sección III de este documento.
- Sin duda alguna, las funcionalidades complementarias atraen al cliente y despiertan interés en los investigadores. Sin embargo, la precisión es un aspecto clave de éxito en los diferentes reconocedores faciales. Los algoritmos encargados de procesar información y construir resultados juegan un papel determinante en la calidad de la herramienta de reconocimiento facial.
- La aplicabilidad del reconocimiento facial tiene diversos campos de uso en el mundo real. Desde el ámbito académico hasta la misma industria, podemos encontrar diversas áreas de aplicación.

## V. HERRAMIENTAS INVOLUCRADAS PARA EL LABORATORIO

Para la elaboración de este artículo se ejecutó un laboratorio en donde fueron utilizados los siguientes elementos:
1. Dispositivo IPAD AIR 2 con el software *Affdex Research* para reconocimiento facial.
2. Audífonos de alta fidelidad marca Sony.
3. Audios asociados a canciones representativas con atributos de alta calidad, para mejorar la experiencia del usuario.
4. Una evaluación de experiencia afectiva de usuario (User Experience - UX) de tipo *emocards* (Tarjetas con connotación emocional) como método de estudio para laboratorios.[12]
5. Unas gafas de realidad virtual VR BOX.

La descripción de las emociones por parte de la persona que experimenta una acción no es una tarea sencilla. Por esta razón en el laboratorio fue clave facilitar al oyente expresar las emociones que percibe. Determinando que el *UX* afectivo más práctico e intuitivo es el de las *emocard*. [13, p. 4]

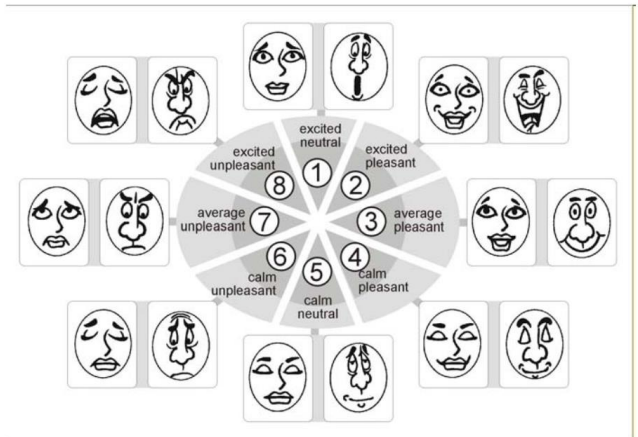

**Figura 13:** Las 8 categorías emocionales y las Emocards. [13, p. 6]

Las *emocard* le permiten al usuario expresar con mayor facilidad y menor ambigüedad sus estados emocionales frente a una determinada vivencia. Este tipo de práctica corresponde al uso de un modelo dimensional de la psicología; en donde se busca acotar las alternativas de selección por parte del usuario, generando una mayor precisión en los resultados.[14, p. 2]

Para el caso de estudio, se utilizaron 2 *emocards* básicas:
Emociones Positivas:
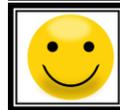
Emociones Negativas:
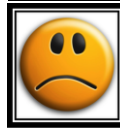

## VI. CASO DE ESTUDIO I

Después de seleccionar 5 personas voluntarias, se les solicitó escuchar atentamente durante 3 minutos cada una de las piezas musicales elegidas.

El procedimiento detallado del caso de estudio fue el siguiente:
1. Se eligieron 2 canciones en las cuales el contraste musical era muy marcado. Para ello se consideró:
   a. Línea musical: Clásica.
   b. Momento histórico de la composición.
   c. Ritmo.
   d. Armonía.
   e. Experiencia personal.
   f. Característica melódica.
2. Se tomó registro fílmico con *Affdex* de cada uno de los oyentes mientras escuchaba cada una de las canciones.
3. Se le aplicó a cada oyente el test UX afectivo después de cada canción.
4. Se realizó una breve entrevista para obtener retroalimentación detallada por parte del oyente.
5. Se confrontaron los resultados fílmicos, el test UX afectivo y las observaciones de la entrevista, para finalmente especificar los resultados.

## VII. RESULTADOS DEL CASO I

Las piezas musicales elegidas para este laboratorio fueron las siguientes:

- *Lacrimosa* de Mozart.
- *Cuarto movimiento de la 9 sinfonía* de Beethoven.

Para cada uno de los voluntarios se obtuvo los siguientes resultados:

Voluntario 1:

Evaluación con *Emocards*:

| 9 S. Beethoven | Lacrimosa – Mozart |
|---|---|

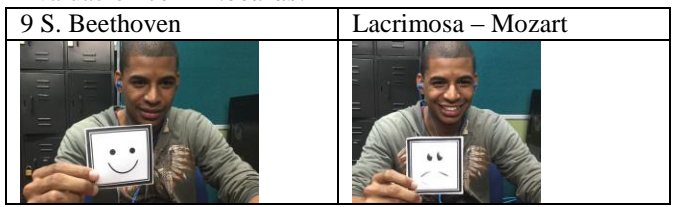

Evaluación con *Affdex Research*:

| 9 S. Beethoven | Lacrimosa – Mozart |
|---|---|

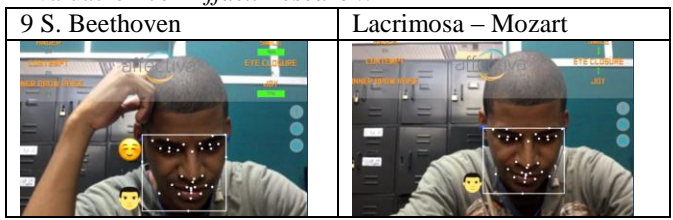

*Affdex* mostró mucha más actividad en la *9 S. Beethoven*, se presentaron algunos intentos de sonrisa. Durante la reproducción de *Lacrimosa*, el oyente se mantuvo muy sereno y *Affdex* no registró actividad relevante en sus gestos.

Comentarios obtenidos en la entrevista:

| 9 S. Beethoven | Lacrimosa – Mozart |
|---|---|
| • Describe un sentimiento de alegría. <br> • Le recuerda algunas etapas de la infancia cuando interpretaba flauta. <br> • Le incita a moverse. | • Se generan recuerdos sobre series de animes de tipo Gore. <br> • Le facilita mantenerse en silencio. |

Voluntario 2:

Evaluación con *Emocards*:

| 9 S. Beethoven | Lacrimosa – Mozart |
|---|---|

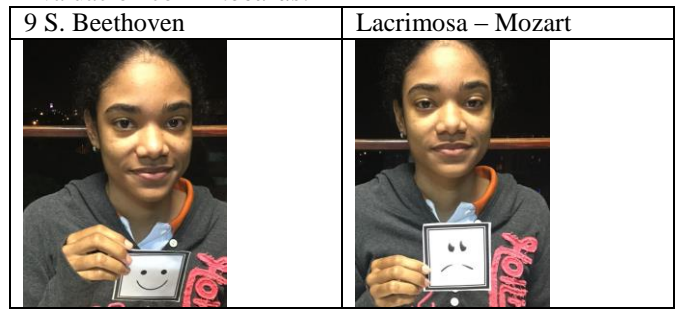

Evaluación con *Affdex Research*:

| 9 S. Beethoven | Lacrimosa – Mozart |
|---|---|

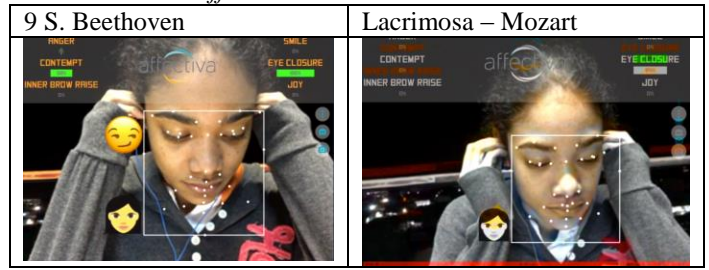

*Affdex* no registró actividad relevante en sus gestos, para ninguna de las 2 piezas musicales. El voluntario involucrado en esta evaluación generó muy poca expresión facial.

Comentarios obtenidos en la entrevista:

| 9 S. Beethoven | Lacrimosa – Mozart |
|---|---|
| • Reconoció la canción y le trajo recuerdos. <br> • Le recuerda su infancia. | • La hace pensar en situaciones problemáticas. <br> • Se imagina que algo puede terminar mal. <br> • Describió el sonido como "Oscuro". |

Voluntario 3:
Evaluación con *Emocards*:

| 9 S. Beethoven | Lacrimosa – Mozart |
|---|---|

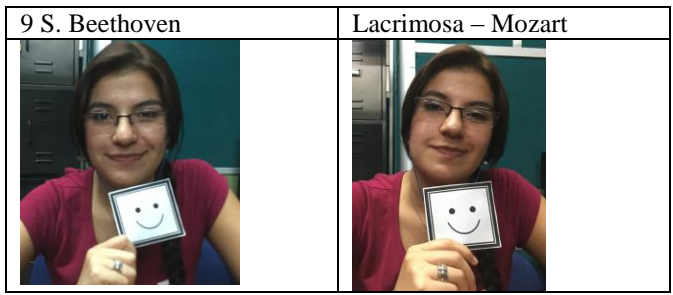

Evaluación con *Affdex Research*:

| 9 S. Beethoven | Lacrimosa – Mozart |
|---|---|

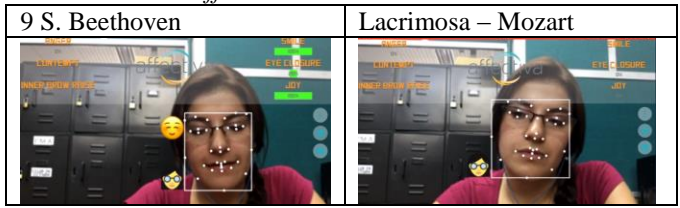

*Affdex* mostró mucha más actividad en la 9 S. Beethoven, en comparación con *Lacrimosa*. El oyente involucrado en esta evaluación, presentaba mucha expresión facial. Durante la reproducción de la *9 S. Beethoven* se presentaron varias sonrisas y algunos movimientos de cabeza.

Comentarios obtenidos en la entrevista:

| 9 S. Beethoven | Lacrimosa – Mozart |
|---|---|
| • Sintió Alegría.<br>• Le pareció un excelente tema para lograr concentración. | • Le pareció una canción equilibrada. |

Voluntario 4:
Evaluación con *Emocards*:

| 9 S. Beethoven | Lacrimosa – Mozart |
|---|---|

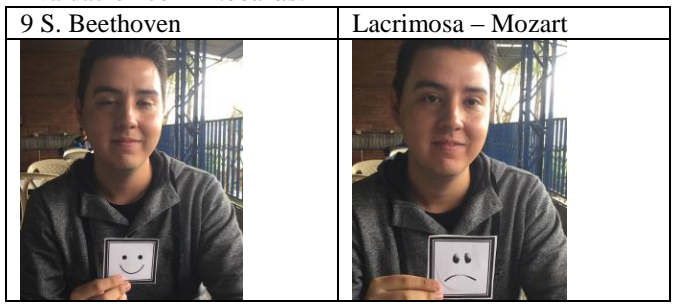

Evaluación con *Affdex Research*:

| 9 S. Beethoven | Lacrimosa – Mozart |
|---|---|

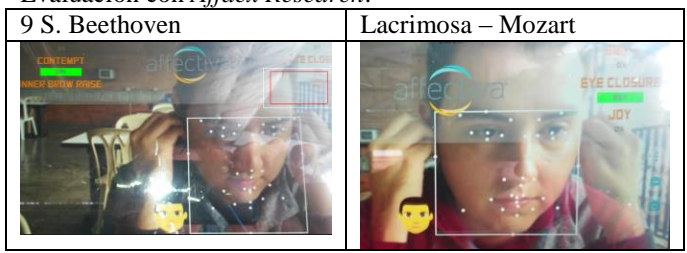

Comentarios obtenidos en la entrevista:

| 9 S. Beethoven | Lacrimosa – Mozart |
|---|---|
| • Le generó un sentimiento de alegría.<br>• Despertó muchos recuerdos de la infancia. | • Percibe inmersión en la soledad.<br>• Se imagina una persona alejándose de todo. |

*Affdex* mostró mucha más actividad en la 9 S. Beethoven, en comparación con *Lacrimosa*. Sin embargo, no se presentó un SMILE de un 100% en *Affedex*. De resaltar que ante un intento notable de sonrisa, el reconocedor facial identificó la emoción de desprecio (Contempt) con un 100% de intensidad.

Voluntario 5.

Evaluación con *Emocards*:

| 9 S. Beethoven | Lacrimosa – Mozart |
|---|---|

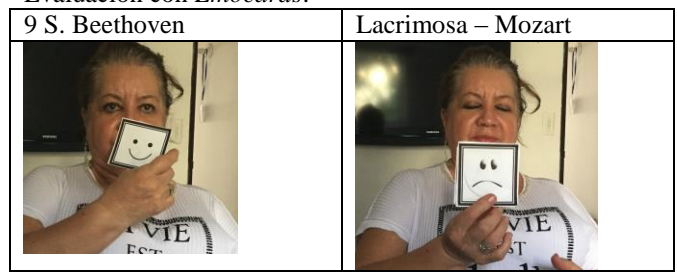

Evaluación con *Affdex Research*:

| 9 S. Beethoven | Lacrimosa – Mozart |
|---|---|

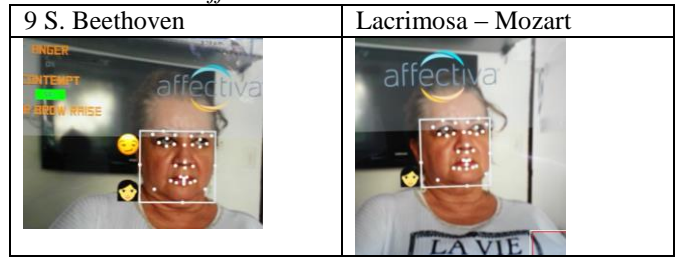

*Affdex* mostró mucha más actividad en la *9 S. Beethoven*, en comparación con *Lacrimosa*. El oyente involucrado en esta evaluación, presentaba expresión facial moderada. Durante la reproducción de la *9 S. Beethoven* se presentaron varias sonrisas y algunos movimientos de cabeza.

Comentarios obtenidos en la entrevista:

| 9 S. Beethoven | Lacrimosa – Mozart |
|---|---|
| • Percibe mucha alegría.<br>• Se refleja en su pie movimiento siguiendo el ritmo. | • Percibe tristeza.<br>• Asoció la canción con sonidos clásicos de iglesia. |

## VIII. CASO DE ESTUDIO II

Para este caso de estudio se incluyó unas gafas de realidad virtual *VR Vox*, y como objetivo, se planteaba evaluar los efectos de la estimulación sonora, a partir de la percepción emocional, que un usuario experimentaba al interactuar con un escenario de realidad virtual.

En cuanto al sonido, los usuarios fueron expuestos a las mismas piezas musicales del caso de estudio presentado en la sección VII de este documento.

Los voluntarios involucrados en este laboratorio, corresponden a un grupo de usuarios totalmente diferentes a los del primer caso de estudio. Esto con el objetivo de evitar distorsión en los resultados, en consideración a que el primer grupo ya conocía gran parte de la dinámica.

Para la parte de realidad virtual de este laboratorio, se consideró utilizar un juego disponible para celulares inteligentes con sistema operativo *android*.

El juego seleccionado fue *House of Terror*, en donde el usuario puede desplazarse por un escenario virtual, con unas características peculiares de misterio y suspenso.

Para este caso de estudio, no se consideró la utilización de programas (software) de reconocimiento facial, en consideración a que las gafas de realidad virtual, no permiten capturar correctamente la expresión facial.

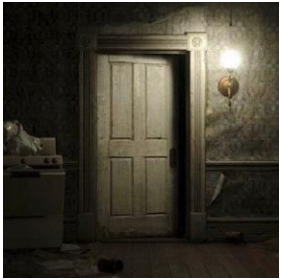

**Figura 14:** Escenario House of Terror.

Después de seleccionar 4 personas, se les solicitó escuchar atentamente cada una de las piezas musicales elegidas, mientras interactuaba con el juego *House of Terror*.

El procedimiento detallado del caso de estudio fue el siguiente:

1. Se seleccionaron las mismas canciones del primer caso de estudio:
   a. *Lacrimosa* de Mozart.
   b. *Cuarto movimiento de la 9 sinfonía* de Beethoven.
2. El usuario jugó alrededor de 10 minutos *House of Terror*. Durante la sesión de juego escucho las canciones descritas en el punto anterior.
3. Se le aplicó a cada jugador la evaluación (test) UX afectivo, después de cada sesión de juego, en donde escuchaba las piezas musicales.
4. Se realizó una breve entrevista para obtener retroalimentación detallada por parte del jugador.
5. Se confrontaron los resultados del *test UX* afectivo y las observaciones de las entrevistas, para finalmente especificar los resultados.

## IX. RESULTADOS DEL CASO II

Para cada uno de los voluntarios se obtuvo los siguientes resultados:

Voluntario 1:

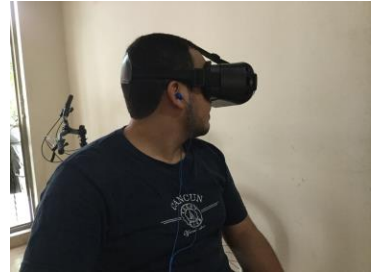

Evaluación con Emocards:

| 9 S. Beethoven | Lacrimosa – Mozart |
|---|---|
| | |

Comentarios obtenidos en la entrevista:

| 9 S. Beethoven | Lacrimosa – Mozart |
|---|---|
| <ul><li>Percibe confianza en el escenario.</li><li>Se motiva a explorar el escenario con mayor libertad.</li><li>Percibe que el ángulo de observación es más alto.</li></ul> | <ul><li>Se siente obligado a recorrer el escenario con más cuidado y lentitud.</li><li>Percibió mayor miedo al jugar.</li><li>Percibió que jugaba agachado, o con un ángulo de observación cercano al suelo.</li></ul> |

Voluntario 2:

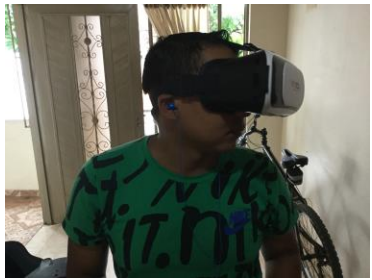

Evaluación con Emocards:

| 9 S. Beethoven | Lacrimosa – Mozart |
|---|---|
| 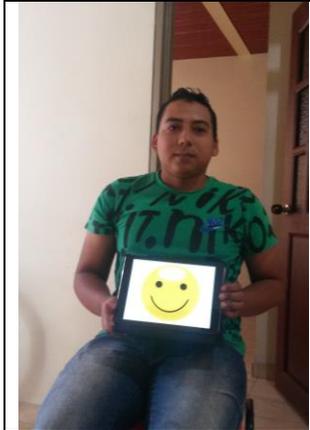 | 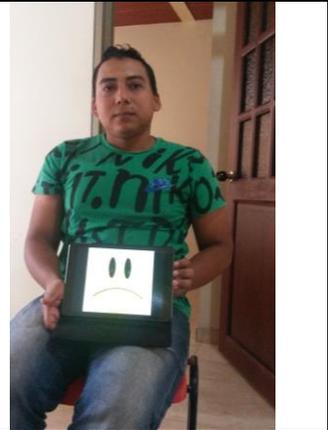 |

Comentarios obtenidos en la entrevista:

| 9 S. Beethoven | Lacrimosa – Mozart |
|---|---|
| - Describe mayor tranquilidad al recorrer el escenario.<br>- Percibió una velocidad de recorrido más alta, por sentimiento de confianza. | - Percibe mayor suspenso al recorrer el escenario de realidad virtual. |

Voluntario 3:

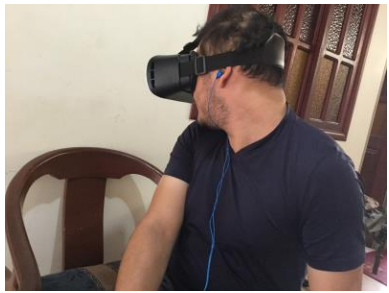

Evaluación con Emocards:

| 9 S. Beethoven | Lacrimosa – Mozart |
|---|---|
| **Sin Calificación** | 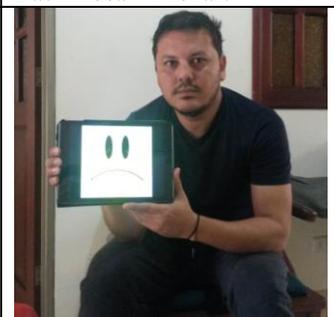 |

El voluntario 3 invitado a este laboratorio, es músico de profesión, y tiene más de 10 años de experiencia como intérprete y compositor.

Comentarios obtenidos en la entrevista:

| 9 S. Beethoven | Lacrimosa – Mozart |
|---|---|
| - Indica que no existe coherencia entre la música y el escenario de realidad virtual.<br>- El voluntario se abstiene de calificar, porque argumenta que ninguna de las opciones del *test* afectivo, se ajusta a la experiencia de usuario. | - Existe coherencia musical con el escenario de realidad virtual.<br>- Percibió suspenso y sentido fúnebre durante la sesión de juego. |

Voluntario 4:

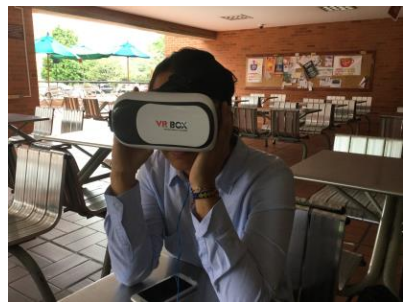

Evaluación con Emocards:

| 9 S. Beethoven | Lacrimosa – Mozart |
|---|---|
| 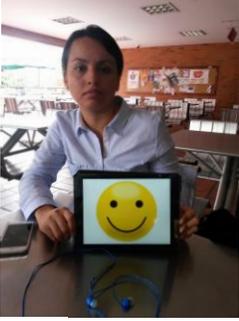 | 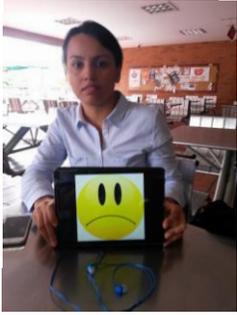 |

Comentarios obtenidos en la entrevista:

| 9 S. Beethoven | Lacrimosa – Mozart |
|---|---|
| • Se percibió mayor seguridad para interactuar con el escenario. <br> • Menor intensidad de sentimiento del miedo. | • Se percibió temor para interactuar con el escenario. <br> • Gran sentimiento de inseguridad y miedo. |

## X. CONCLUSIONES RELATIVAS A LOS CASOS DE ESTUDIO

Después de realizar las diferentes sesiones con cada uno de los participantes en los dos laboratorios, se plantean las siguientes conclusiones.

Caso I:

- En general *Affdex* registró una mayor cantidad de actividad en sus mediciones emocionales durante la reproducción de la *9. S. De Beethoven*.
- En general *Affdex* registró muy poca actividad en sus mediciones emocionales durante la reproducción de *Lacrimosa* de Mozart.
- Existe una relación entre los resultados obtenidos con el test UX afectivo y los resultados obtenidos por la aplicación de reconocimiento facial.
- Falta precisión en el reconocimiento de algunos gestos particulares por parte de la herramienta de reconocimiento facial. Para el caso del voluntario 4, un gesto neutral fue identificado como un gesto de desprecio. De igual manera varias sonrisas no fueron identificadas.
- El ambiente en donde se desarrolla este tipo de experimentos es un factor importante. Para la mayoría de las sesiones se seleccionó un sitio tranquilo y despejado. Adicionalmente se dieron instrucciones de concentración al oyente, y sólo hasta el final del ejercicio se les explicó el propósito del laboratorio. Por lo tanto, el oyente no era consciente del análisis que se estaba realizando.
- No todos los oyentes generaron la misma intensidad en su expresión facial. El voluntario 2 generó la menor cantidad de expresión facial. El análisis de este voluntario se dio fundamentalmente a través de las *emocards* y de la entrevista.
- Sólo 1 voluntario de 5 en total, no asoció una emoción negativa a través de *emocard* a la pieza musical *Réquiem* de Mozart.
- La música esta asociada con recuerdos. Los 5 participantes de este laboratorio asociaron la *9. S. Beethoven* con algún recuerdo de sus vidas.
- Se logró evidenciar una respuesta de gestos faciales frente a la actividad de escuchar música. Sin embargo, no en todos los voluntarios se presentaron con la misma intensidad. Y en muchos de los casos, la herramienta presentó mediciones poco precisas.
- Los resultados de este ejercicio dan lugar a buscar otro tipo de señales que puedan ser medidas con mayor precisión. Y así, seguir trabajando en plantear un modelo explicativo sobre el efecto de la música en las emociones y el reconocimiento de estas mismas desde la computación.

Caso II:

- Generalmente el jugador intenta relacionar el sonido con el contenido gráfico del ambiente virtual; y plantea un calificativo de coherencia.
- Las características del sonido, desde el punto de vista de atributos de la música, puede afectar la interpretación emocional del jugador que interactúa con un ambiente de realidad virtual.
- El componente visual genera un efecto emocional en el jugador a través de la inmersión. De igual manera, el componente de sonido puede intensificar o debilitar una misma emoción.
- Una apropiada relación entre el ambiente de realidad virtual y el contexto de sonido, genera un efecto de distracción lógico y apropiado sobre la experiencia de usuario. Tal efecto de distracción ha sido utilizado incluso para tratar algunos procedimientos clínicos en pacientes, logrando reducir considerablemente la percepción del dolor [15].
- El sonido puede fortalecer o debilitar el efecto emocional que un escenario de realidad virtual pretende generar. Es importante, estudiar y establecer adecuadamente escenarios de realidad virtual con un contexto de sonido apropiado, para así lograr una mejor inmersión por parte del usuario.

# XI. CONCLUSIONES RELATIVAS AL EFECTO DE LA MÚSICA SOBRE LA REALIDAD VIRTUAL

Desde la perspectiva de la música la gran mayoría de los compositores utilizan sus emociones y sentimientos como fuente de inspiración para desarrollar sus obras musicales.

Este material emocional que un compositor transfiere a su obra musical, generalmente es percibido por un oyente, quien a su vez dependiendo de su estado psicológico y otros factores de contexto responde emocionalmente.

La música tiene un efecto emocional, y el ser humano lo confirma de manera empírica. Las señales faciales pueden ser medidas por herramientas de reconocimiento, y el efecto generado por la música se detecta, aunque no siempre con la misma precisión.

Existen algunos marcos de trabajo (frameworks) desarrollados al momento, que permiten procesar canciones y extraer características de audio. Estos *frameworks* cuentan con algoritmos que permiten ubicar las canciones dentro de ciertas categorías. Sin embargo, algunas características de la música, como es el caso de la tonalidad; no siempre se identifica apropiadamente. [14, p. 3]

La inmersión es un aspecto clave en la realidad virtual, y aunque el componente gráfico es un elemento con gran desarrollo en la actualidad, no podemos dejar de lado el sonido. Y para ser más específico, el sonido desde su expresión musical. A través del segundo caso de estudio, fue posible evidenciar que existe un efecto del sonido sobre una experiencia de realidad virtual.

El sonido nos puede transportar a muchos lugares. La connotación emocional a través de la música, resulta clave en un sistema de realidad virtual efectivo; entendiendo que se requiere una coherencia musical en relación al escenario de realidad virtual.

Probablemente para las ciencias computacionales, resulte un reto modelar el algoritmo con el que un compositor crea música. O quizás modelar un sistema con la capacidad de clasificar emocionalmente una canción, simulando una evaluación humana. Y aún más interesante modelar un sistema dinámico, capaz de improvisar música según el estado de ánimo detectado en una persona; creando un sistema de realidad virtual totalmente dinámico en su generación de sonido, y acorde a los escenarios presentados durante la experiencia de usuario.